\def\corr #1 { #1}
\documentclass[epsf,psfig]{aa}

\begin{document}

\thesaurus{02         
               (02.13.2)  
               (02.20.1)  
               (03.13.4)  
               (06.13.1)  
               (09.13.1)}  
\title{{\bf Ambipolar Filamentation of Turbulent
Magnetic Fields : \\ A numerical simulation.}}

\author{M. Franqueira\inst{1,2} \and M. Tagger\inst{1} \and  A. I. 
G\'omez de Castro\inst{2} }

\offprints{mf@orion.mat.ucm.es}

\institute{Service d'Astrophysique (CNRS URA 2052), CEA/DSM/DAPNIA, CE
Saclay, 91191 Gif sur Yvette, France
\and Instituto de Astronom\a'{\i}a y Geodesia, CSIC-UCM, Madrid, Spain\\}

\date{\number\day.\number\month.\number\year}

\titlerunning{Numerical Simulation of Ambipolar Filamentation}
\authorrunning{M. Franqueira {\it et al.}}

  \maketitle

\begin{abstract}

We present the results of a 2-D, two fluid (ions and neutrals) simulation
of the ambipolar filamentation process, in which a magnetized, weakly
ionized plasma is stirred by turbulence in the ambipolar frequency
range.  The higher turbulent velocity of the neutrals in the most
ionized regions gives rise to a non-linear force driving them out of
these regions, so that the initial ionization inhomogeneities are
strongly amplified.  This effect, the ambipolar filamentation, causes
the ions and the magnetic flux to condense and separate from the neutrals,
resulting in a filamentary structure.

\keywords{MHD -- Turbulence -- Methods: numerical -- Sun: magnetic 
fields -- ISM:
magnetic fields }

\end{abstract}

\vskip 0.5cm

\section{Introduction}

Magnetic fields contribute to the dynamical behavior of ionized astrophysical
fluids such as those in the upper solar and stellar atmospheres, the
interstellar medium  and star-forming regions. Their
influence is carried out
by hydromagnetic waves which efficiently propagate
perturbations, ensure a turbulent pressure or may even
cause the development of instabilities (\cite{arons}).

However, Kulsrud \& Pearce (\cite{kulsrud}) showed that in the
magnetized and weakly ionized interstellar medium hydromagnetic waves
are heavily damped in a frequency range (and thus scale) associated with
ambipolar diffusion.  At low frequency the neutrals are well coupled to
the ions (which are tied to the magnetic field lines) and hydromagnetic
waves propagate at the Alfv\'en speed defined by the total inertia
(given by ions+neutrals).  At high frequency neutrals and ions are
totally decoupled, and Alfv\'en waves involve only the ions, which
define a larger Alfv\'en velocity.  In the intermediate range (the
`ambipolar range', between the ion-neutral and neutral-ion collision
frequencies $\nu_{in}$ and $\nu_{ni}$) the neutrals are imperfectly
coupled to the ions; this results in a drag which strongly damps the
waves \corr{(see also \cite{mcivor}, for a description of MHD turbulence
in the partly ionized interstellar medium)}.

The non-linear evolution of this process can cause an {\em ambipolar
filamentation} of the magnetic field when a magnetized and weakly
ionized plasma is stirred by hydromagnetic turbulence in the ambipolar
range (\cite{tagger}).  If such a plasma presents small variations in
the ionization fraction ($Z=\rho_{i}/\rho_{n}$), the turbulent velocity
of the neutrals is higher in the most ionized regions, since they are
better coupled to the ions.  This gives rise to a force (given by the
average of the $v.\nabla v$ term) driving the neutrals out of the most
ionized regions.  By reaction the ions and the magnetic flux are
compressed in these regions, so that the initial ionization
inhomogeneities are strongly amplified.  As a consequence a
concentration of the flux tubes is expected to occur, producing a
filamentary structure, so that turbulent energy would be converted into
magnetic energy associated with the concentration of the magnetic field.
Tagger {\it et al.} (1995) provided only order of magnitude estimates of
the expected amplification of the ionization fraction.  In this work we
present a fully consistent 2-D non-linear numerical simulation of the
mechanism in order to test its efficiency.

The non-linear analysis is a fundamental
tool to study the physics in certain astrophysical environments, such as
molecular clouds, where the observed amplitudes of the turbulent velocities
are comparable with the mean field velocities.
The ambipolar filamentation mechanism might help to explain some well
known problems arising in magnetized, partially ionized astrophysical
plasmas. One of them is related with the observations of turbulence in
molecular
clouds.  Observations show a filamentary structure, and strong
supersonic motions resulting in turbulent and magnetic energies
in approximate equipartition, i.e., much larger than the
thermal energy (\cite{myers}).  The ambipolar filamentation
mechanism would concentrate the magnetic field in intense flux ropes
surrounded by essentially neutral clouds.

Another possible application relates to the fibrilled structure observed
in the magnetic field emerging from the solar photosphere, organized in
very narrow flux tubes.  The ambipolar filamentation mechanism might
provide an explanation for the spicules emerging from the photosphere:
let us consider magnetic field lines raising from the photosphere.  Then
an Alfv\'en wave of a given frequency, produced in the photosphere and
initially below the local ambipolar frequency range, will propagate upward
along
the field lines and reach at high altitudes a plasma of much lower
density, i.e., lower collision frequencies.  It will thus be damped by
ambipolar effects and can expel the neutrals from the most ionized flux
tubes, concentrating the magnetic flux in narrow tubes where strong
vertical motions can be expected.  This would occur together with the
mechanism discussed by De Pontieu \& Haerendel (\cite{depon}).  These 
prospects will be
discussed in more detail in the last section of this work.

We have carried out numerical simulations in which a weakly ionized and
magnetized gas inside a cartesian box is submitted to a high amplitude
oscillation  emitted from one of its sides.  The perturbation 
propagates inside
the box as an Alfv\'en wave with a frequency chosen to be in the
ambipolar range, so that it will be strongly damped.  In
Section  2 we describe the dynamical equations that govern the evolution
of a two fluid gas, together with the numerical code and the  boundary
conditions used to
solve them.  We also discuss the numerical constraints present in our
simulations.  The results from the numerical experiments are presented
in Section  3 and discussed in the context of the problems cited above in
Section  4.

\section{The Numerical Code}

The magnetohydrodynamics (MHD) equations describing a two fluid (ions
and neutrals) system are (\cite{langer}):

\begin{equation}
\label{eq1}\frac{\partial \rho_i }{\partial t}+\nabla
\left( \rho_i \vec{v}_i\right)=0
\end{equation}

\begin{equation}
\label{eq2}\frac{\partial \rho_n }{\partial t}+\nabla
\left( \rho_n \vec{v}_n\right)=0
\end{equation}

\begin{equation}
\label{eq3}\frac{\partial \vec{v}_i}{\partial t}+\left(
\vec{v}_i.\nabla \right) \vec{v}_i=-\frac{\nabla
p_i}{\rho_i} -
\vec{g}+(\nabla \times \vec{B})\times \vec{B%
}+\mu \rho_n(\vec{v}_n-\vec{v}_i)
\end{equation}

\begin{equation}
\label{eq4}\frac{\partial \vec{v}_n}{\partial t}+\left(
\vec{v}_n.\nabla \right) \vec{v}_n=-\frac{\nabla
p_n}{\rho_n} -
\vec{g}+\mu \rho_i(\vec{v}_i-\vec{v}_n)
\end{equation}

\begin{equation}
\label{eq5}\frac{\partial \vec{B}}{\partial t}=\nabla \times
\left( \vec{v}_i\times \vec{B}\right)\ .
\end{equation}

For simplicity we assume an isothermal equation of state:

\begin{equation}
\label{eq6}
p_i=\rho_i c_s^2 ,
\end{equation}

\begin{equation}
\label{eq7}p_n=\rho_n c_s^2 ,
\end{equation}

\begin{figure}[htb*]
\setlength{\unitlength}{1.0cm}
\begin{picture}(8,10.5)
\includegraphics{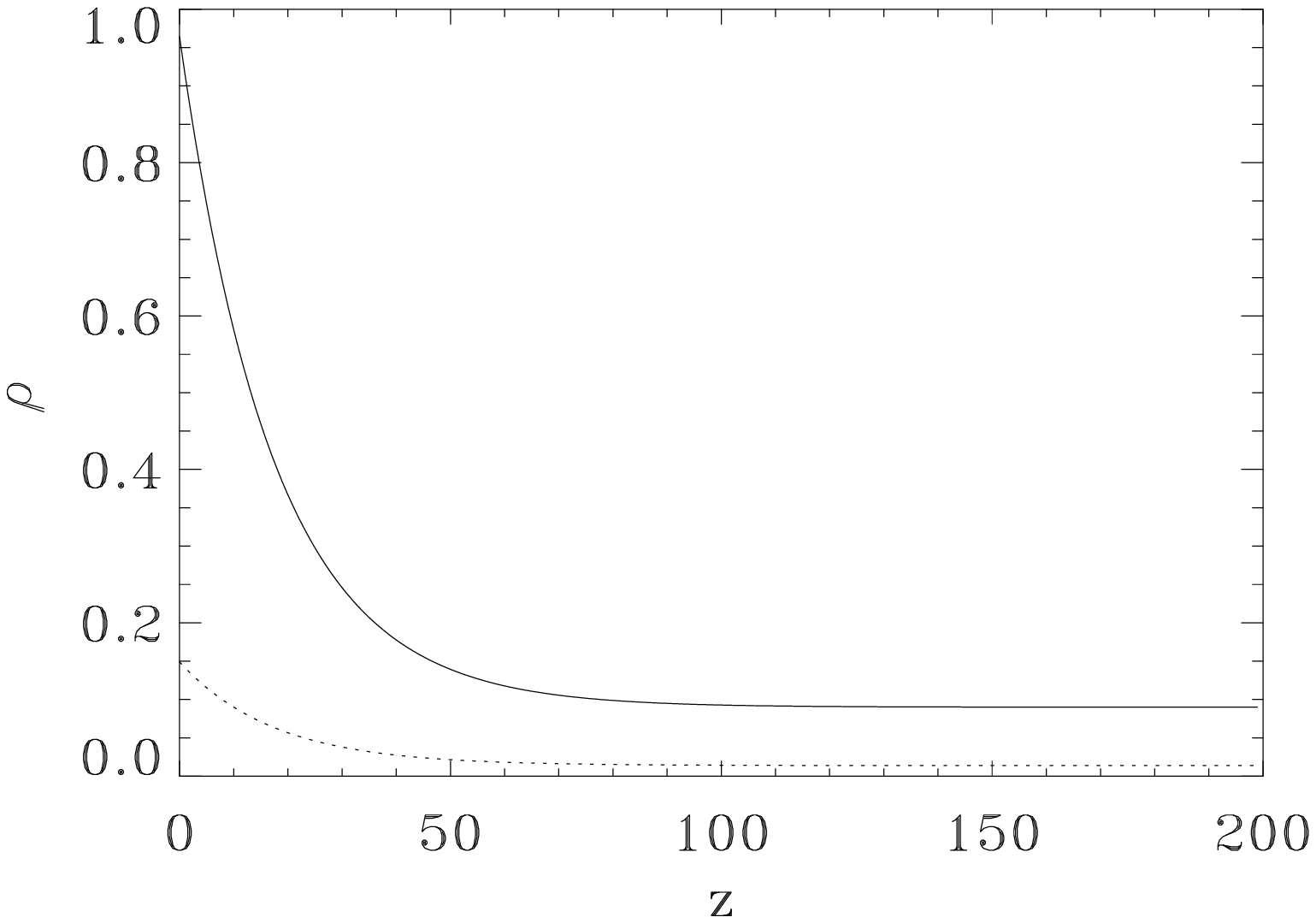}
\includegraphics{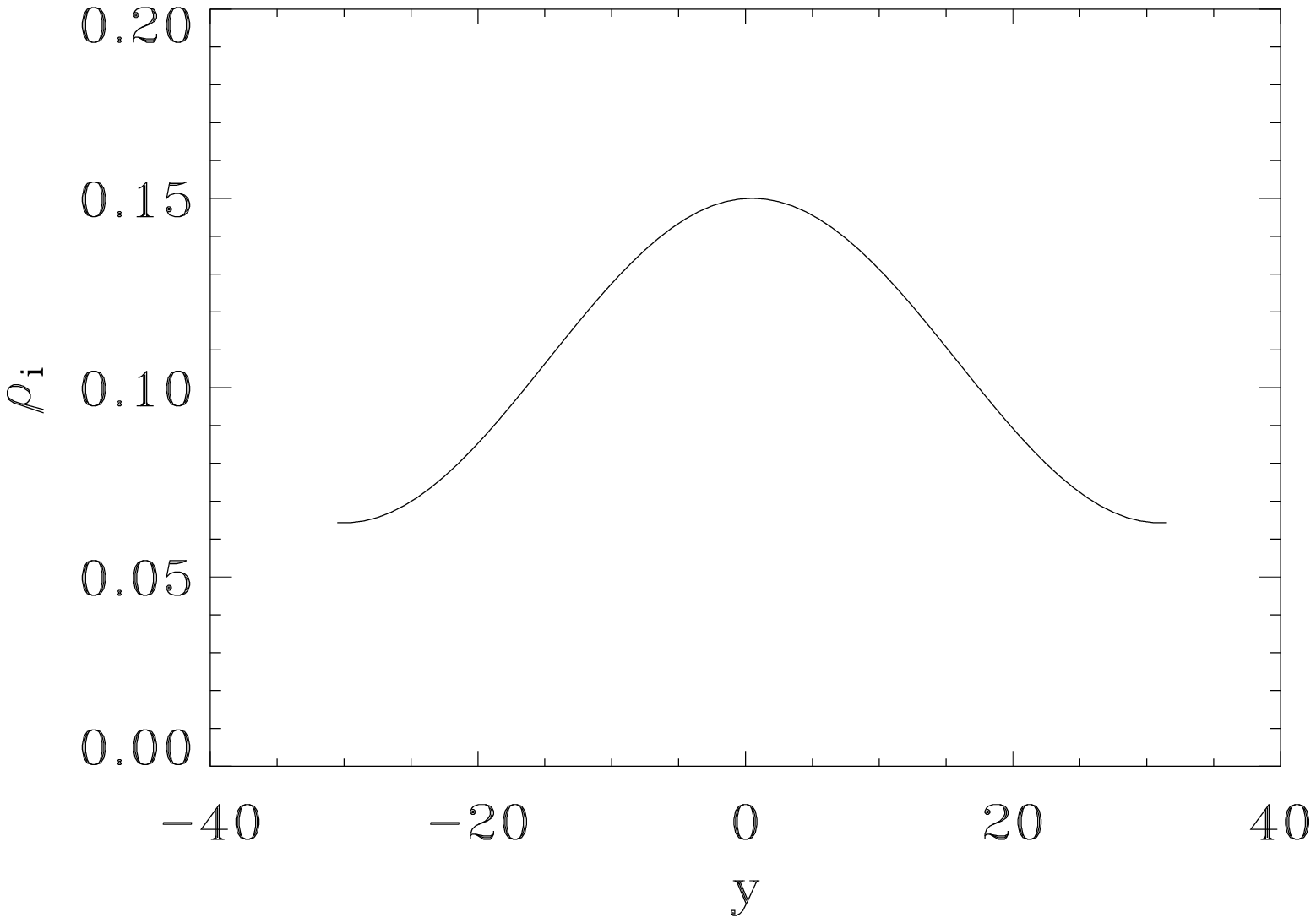}
\end{picture}
  \caption{ Top: Initial ions (dotted line) and neutrals (solid line) density
profiles in the $z$ direction at $y=0$. Bottom: Initial ions density
profile in the $y$ direction at $z=0$. We use the grid zone
indices to scale the horizontal ($y$) and vertical ($z$) directions in all
figures. See Section 3 for an explanation of the parametrization used.}
\label{1f}
\end{figure}

\noindent
where $\rho$, $v$ and $p$ are, respectively, the density, velocity and
partial pressure of the ions (with subscript i) and neutrals (with
subscript n), $g$ is the gravity, $\mu$ is a constant such that
$\nu_{in}=\mu\rho_{n}$ and $\nu_{ni}=\mu\rho_{i}$ are the ion-neutral
and neutral-ion collision frequencies, and $c_{s}$ is the sound velocity
(assumed the same for ions and neutrals).  We assume that ionization and
recombination occur on a longer time scale than the one we consider.
This should of course be checked for applications to specific
astrophysical situations.

We have also checked that in these conditions the characteristics of the
problems in which we are interested, namely the high electron densities
in the case of solar spicules and the large spatial dimensions of
molecular clouds, allow us to use these simplified two fluid MHD
equations instead of the full set of the three fluids (electrons, ions
and neutrals) equations which describe the dynamics of a weakly ionized
gas.

In order to allow for the long time scales considered (hundreds of
Alfv\'en times), associated with numerical constraints giving a short
time step, we simplify the problem by making it two-dimensional: all
quantities are $x$-invariant, and only the perturbed current has a
component along $x$.  Therefore all quantities depend only on $z$ (the
vertical direction) and $y$ (the horizontal one) on a cartesian grid.
In this geometry the distinction between shear-Alfv\'en and magnetosonic
waves disappears, but waves propagating along $z$ retain the properties
of the usual Alfv\'en waves, that is, they twist the field lines and to
lowest order they are not compressional.

The numerical code used in our simulations is based on the same general
methods as the ZEUS-2D code (\cite{stonea}).  It solves the MHD
equations on a staggered mesh using the method of finite differences
with a time explicit, operator split scheme.  Densities and pressures
are zone centered quantities while the velocity components are centered
at their corresponding zone faces.  As in ZEUS-2D the solution procedure
is arranged in two main steps, first taking into account the source
terms in the right-hand side of the equations, and then solving for the
advection terms in the left-hand side.  The main difference with ZEUS-2D
is the treatment of the gravitational and magnetic terms.  An initial
state of hydrostatic equilibrium is assumed, by fixing the vertical
density and pressure profiles, from which we derive a value for the
gravity; this results in an ad-hoc gravity profile leaving us some
freedom to optimize the density profiles, limiting the numerical
difficulties associated with the emission of the wave (see below).  The
2-D geometry allows us to describe the magnetic field in eqs.
(\ref{eq3}) and (\ref{eq5}), in terms of the flux function $\psi$ as
follows:

\begin{equation}
\label{eqpsi}\vec{B}=B_0\vec{e_z}+\vec{e_x}\times \vec\nabla \psi
\end{equation}
so that the induction equation, eq. (\ref{eq5}) becomes:
\begin{equation}
\label{eq9}\frac{\partial}{\partial t}\psi+
\vec{v}_i.\nabla \psi=-v_{y}B_{0}\ .
\end{equation}

The momentum and induction equations are thus first solved without the
advection terms, using time explicit operator split schemes.  The MOC-CT
algorithms included in ZEUS-2D (\cite{stoneb}) are not used to evolve
the magnetic terms, since they are not required by this simple problem.
Tests were carried out to verify the correct propagation of the waves.

The second step in the numerical code, still following ZEUS-2D, solves
finite difference versions of the integral equations coming from
the advection terms in eqs.  (\ref{eq1}), (\ref{eq2}), (\ref{eq3}) and
(\ref{eq4}):

\begin{equation}
\label{eq10}\frac d{dt}\int_V\rho_i \ dV=-\int_{dV}\rho_i \vec{v_i}.
\vec{\ dS}
\end{equation}
\begin{equation}
\label{eq11}\frac d{dt}\int_V\rho_n \ dV=-\int_{dV}\rho_n \vec{v_n}.
\vec{\ dS}
\end{equation}
\begin{equation}
\label{eq12}\frac d{dt}\int_V\rho_i \ \vec{v_i}dV=-\int_{dV}\rho_i
\vec{v_i}\vec{v_i}.\vec{\ dS}
\end{equation}
\begin{equation}
\label{eq13}\frac d{dt}\int_V\rho_n \ \vec{v_n}dV=-\int_{dV}\rho_n
\vec{v_n}\vec{v_n}.\vec{\ dS}
\end{equation}

\noindent
which allows us to obtain a conservative scheme by computing the fluxes
of the advected quantity ($\rho_i$, $\rho_n$, $\rho_i \vec{v_i}$,
$\rho_n \vec{v_n}$ respectively in the equations above) at every
interface on the grid and using the same flux to update adjacent zones.
To solve the problem of calculating values of the advected quantities on
the grid interfaces maintaining numerical stability (\cite{stonea}),
we used the second order van Leer interpolation method (\cite{vanleer}).
This method is fast and accurate enough for the
requirements of our simulations.  As in ZEUS-2D the two-dimensional
advection problem is simplified by using directional splitting (\cite{strang}),
that means using two one-dimensional advection steps to construct
the full solution.

The boundary conditions used to solve the MHD equations are determined
by the physics of the phenomenon we are studying.  We use periodic
boundary conditions in $y$ (note that with the definition of $\psi$ in
equation (\ref{eqpsi}), the periodicity of $\psi$ ensures conservation
of the total vertical magnetic flux through the simulation zone).  At
$z=0$ we launch an Alfv\'en wave by giving the whole fluid (neutrals,
ions and magnetic field lines) a motion in $y$ which, in this first
numerical test of the mechanism, will be limited to a single periodic
oscillation:
\[v_{y}(y, z=0)=v_{t}\cos(\omega t)\ .\]
This will thus
propagate upward as an Alfv\'en wave.  At the opposite boundary
($z_{max}$) we impose reflective boundary conditions; they have no real
effect since the total length in $z$ is chosen such that the waves are
heavily damped before they reach that point.  At the moment, in order to
isolate the effect we want to study, we impose that there is no flux of
matter from $z=0$. In this manner we ensure that the variations in the
ion or neutral densities do not result from inflow of matter at the
lower boundary. On the other hand this results in severe constraints on
the code because the wave pressure pushes matter upward, resulting in
very low densities at the first grid point. This turns out to be the
most stringent limit on the parameters we can use ({\it e.g.} the
perturbed velocity $v_{t}$).

The simulations have other numerical constraints.  To ensure that our
calculations are fully consistent the dimensions of our spatial grid
must be large enough to allow the ion-neutral interaction, that is,
larger than the ion-neutral (neutral-ion) mean free path, $l= v/\nu$,
where $\nu$ is the collision frequency.  For the characteristic values
used in the simulations the grid is more than 10 times the maximum mean
free path for the neutrals,
which is enough for the necessary ion-neutral interaction.

\begin{figure*}[htb*]
\setlength{\unitlength}{1.0cm}
\begin{picture}(8,7)

\includegraphics{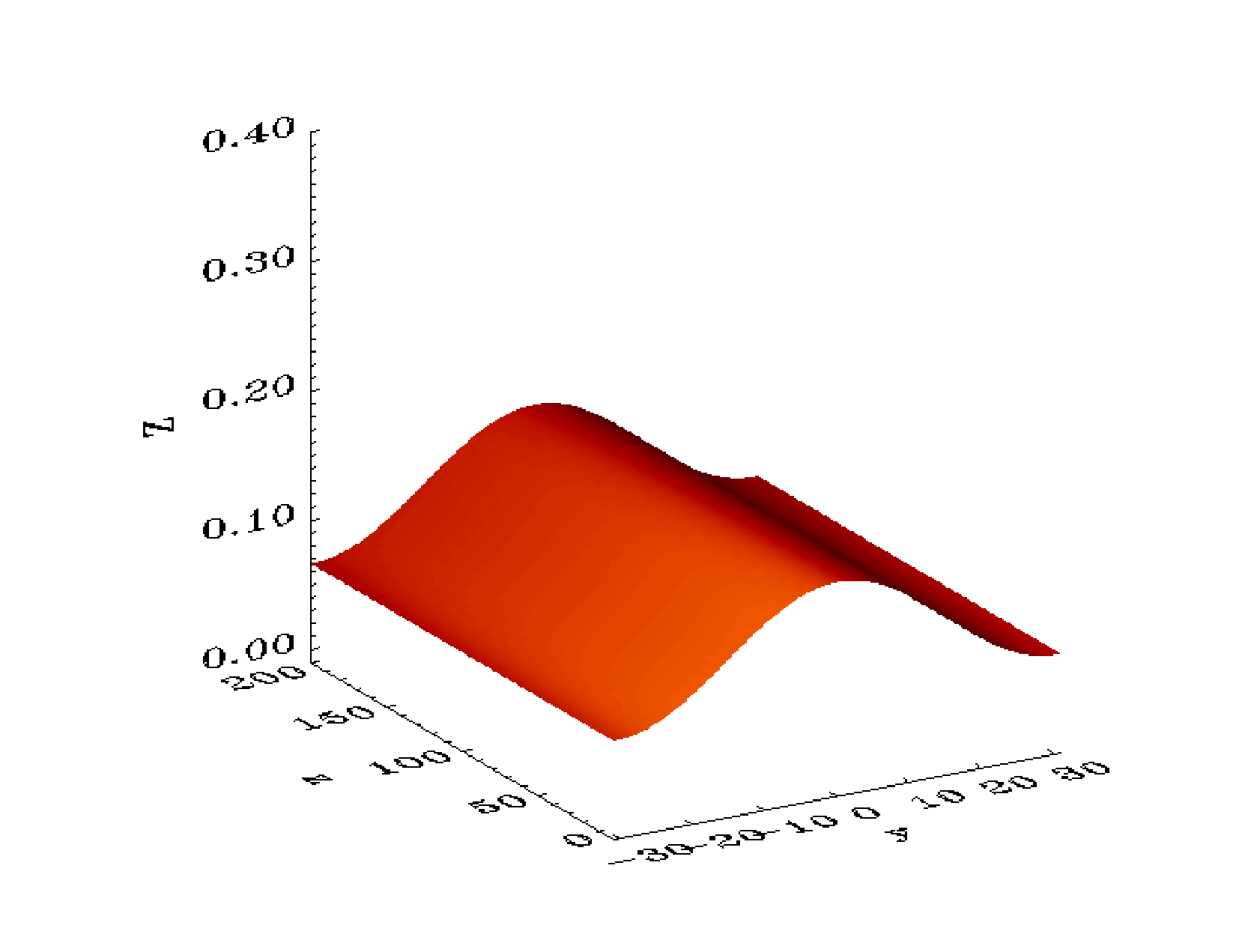}
\includegraphics{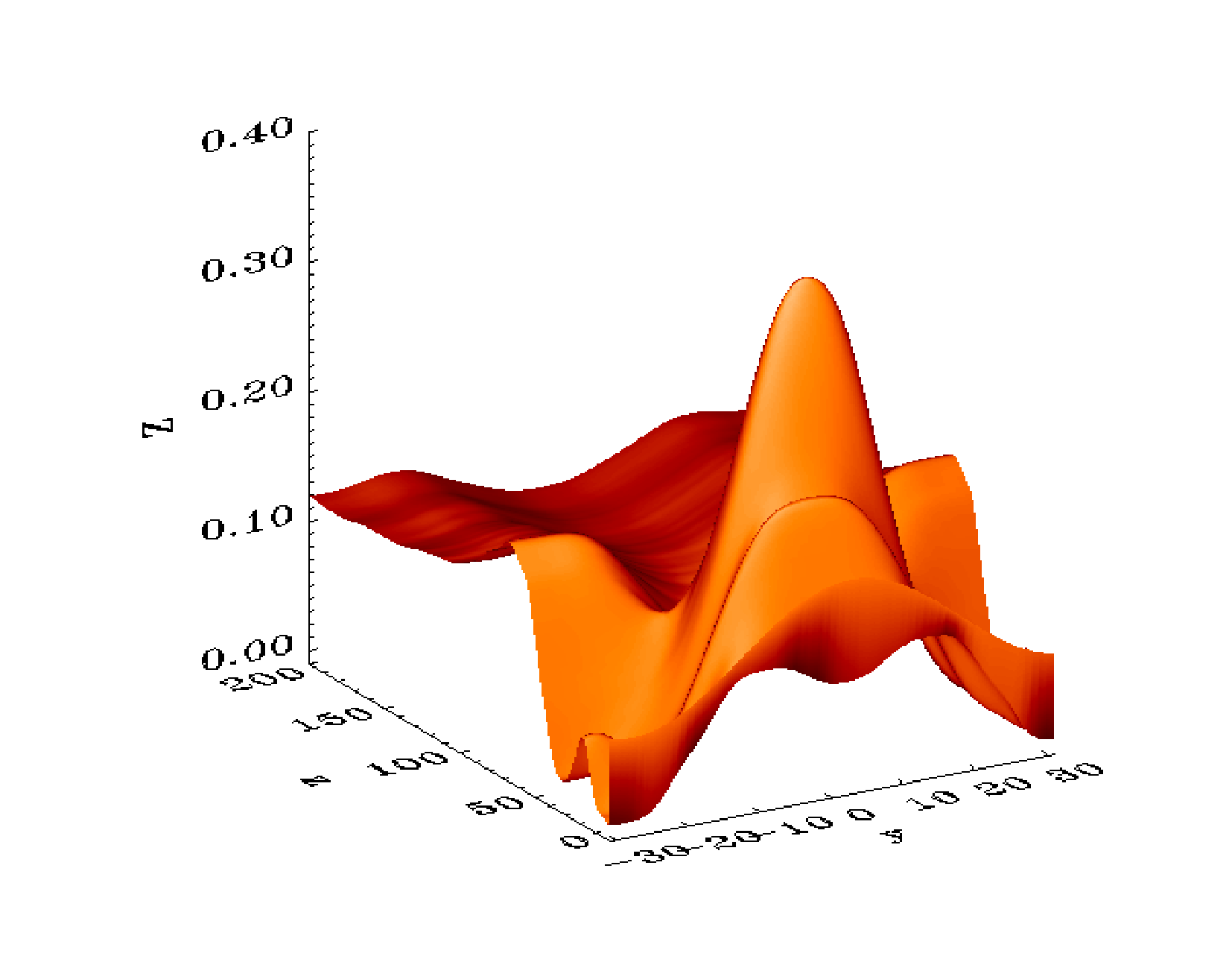}

\end{picture}
  \caption{ Left: Initial  2D distribution of the ionization fraction
$Z=\rho_i/\rho_n$ in the calculation grid. Right: Ionization fraction after
$6.7 \times 10^3$
Alfv\'en times. The contrast along $y$ is strongly enhanced at the
altitude where the wave is damped.}
\label{2f}
\end{figure*}

We also wish to launch the wave in a manner such that it is damped not
right in its emission zone (near $z=0$, where we create it) but away
from it, so that we can clearly identify the processes obtained.  We do
this by imposing a strong vertical density gradient.  Thus the density
at $z=0$ can be taken such that the wave initially propagates well
($\omega\ll\nu_{ni}=\mu\rho_{i}$), but later reaches an altitude where
$\omega\sim\nu_{ni}$ so that it is strongly damped and ambipolar
processes can act.  At higher altitude we maintain a small but constant
density.  Finally, in order to initiate the filamentation process we
need a transverse density profile: the neutral density is initially
independent of $y$, but the ion density has a horizontal profile, so
that the ionization fraction is higher along the central field line
($y=0$).  The magnetic field is chosen to ensure MHD equilibrium in the
horizontal direction, that is:
\begin{equation}
\label{peq}\frac{\partial}{\partial y}\left(p_i+\frac{B^2}{8\pi}\right)=0\ .
\end{equation}
  Thus we expect that ions in the central flux tube will
be compressed, and the neutrals expelled, by the ambipolar
filamentation process.

On the other hand, in order to save computation time we must take the
largest time step that guarantees the numerical stability for the set of
difference equations.  Our time explicit code has to satisfy a
Courant-Friedrichs-Lewy (CFL) stability condition,

\begin{equation}
\label{eq14}\Delta t \leq \frac {\Delta x}{v_{max} },
\end{equation}

\noindent
derived applying a von Neumann stability analysis (\cite{rich}).
Physically, this condition sets the higher limit of the distance
that information can travel in one time step (waves or fluid motion) to
be the minimum size of the discrete elements or 'zones' in the
spatial grid.  For multidimensional systems, a suitable time step
limit is the smallest of all one dimensional CFL conditions in each
coordinate direction (\cite{stonea}). This gives strong
constraints since (a) we need to consider the highest Alfv\'en
velocity, obtained where the density is lowest (b) the CFL condition
involves waves with the shortest wavelengths, {\it i.e.} high
frequencies. At high frequency, as explained in the introduction, the
relevant Alfv\'en velocity involves only the ion density. Therefore compared
with the Alfven velocity at $z=0$ the one used for the CFL criterion is
higher by the density ratio $(\rho_{i,max}/\rho_{i,min})^{1/2}$ along the
vertical density profile, and by the inverse
of the ionization ratio $\rho_{n}/\rho_{i}$. These two quantities are
large and force us to use very short time steps, typically less than
$10^{-2}$ of the Alfv\'en time through a grid cell at $z=0$.

Let us now consider the time scales associated with the mechanism of
ambipolar filamentation. Three time-scales appear.
The first one is the period of the wave. The second one involves the
response of the fluid to the wave pressure; in the basic
mechanism of ambipolar filamentation, the neutrals feel a turbulent
pressure:
\begin{equation}
     p_{Turb}=\frac{1}{2}\rho_{n}\left<v_{n}^2\right>
     \label{pturb}
\end{equation}
(where the angular brackets mean a time average), which has a gradient
along $y$ associated with the gradient of $\rho_{i}$; they respond by a
pressure perturbation to re-establish equilibrium.  This is established
after a time $\sim L/c_s$, where $L$ is the scale of the horizontal
ionization inhomogeneity.

\begin{figure*}[htb*]
\setlength{\unitlength}{1.0cm}
\begin{picture}(8,9)
\includegraphics{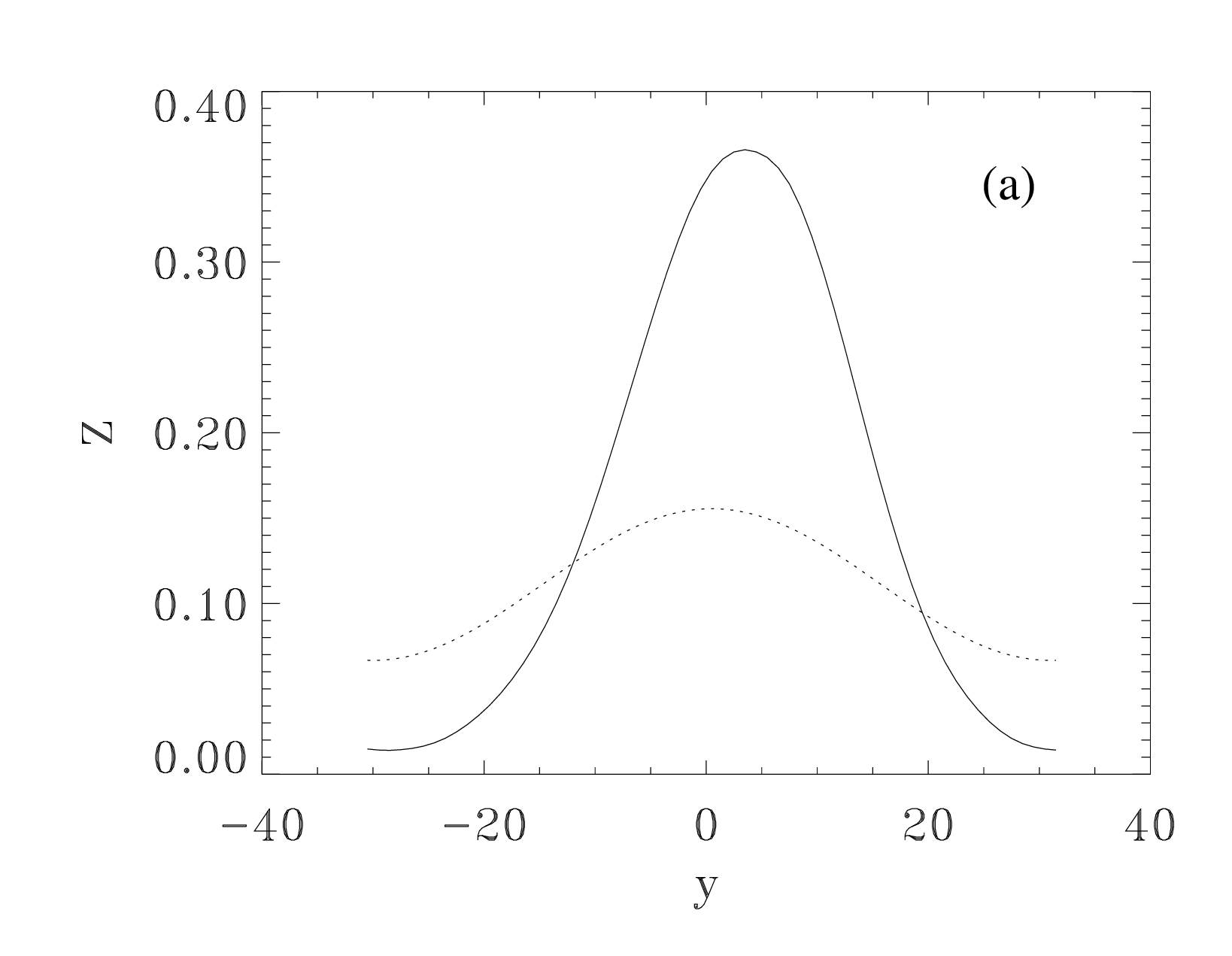}
\includegraphics{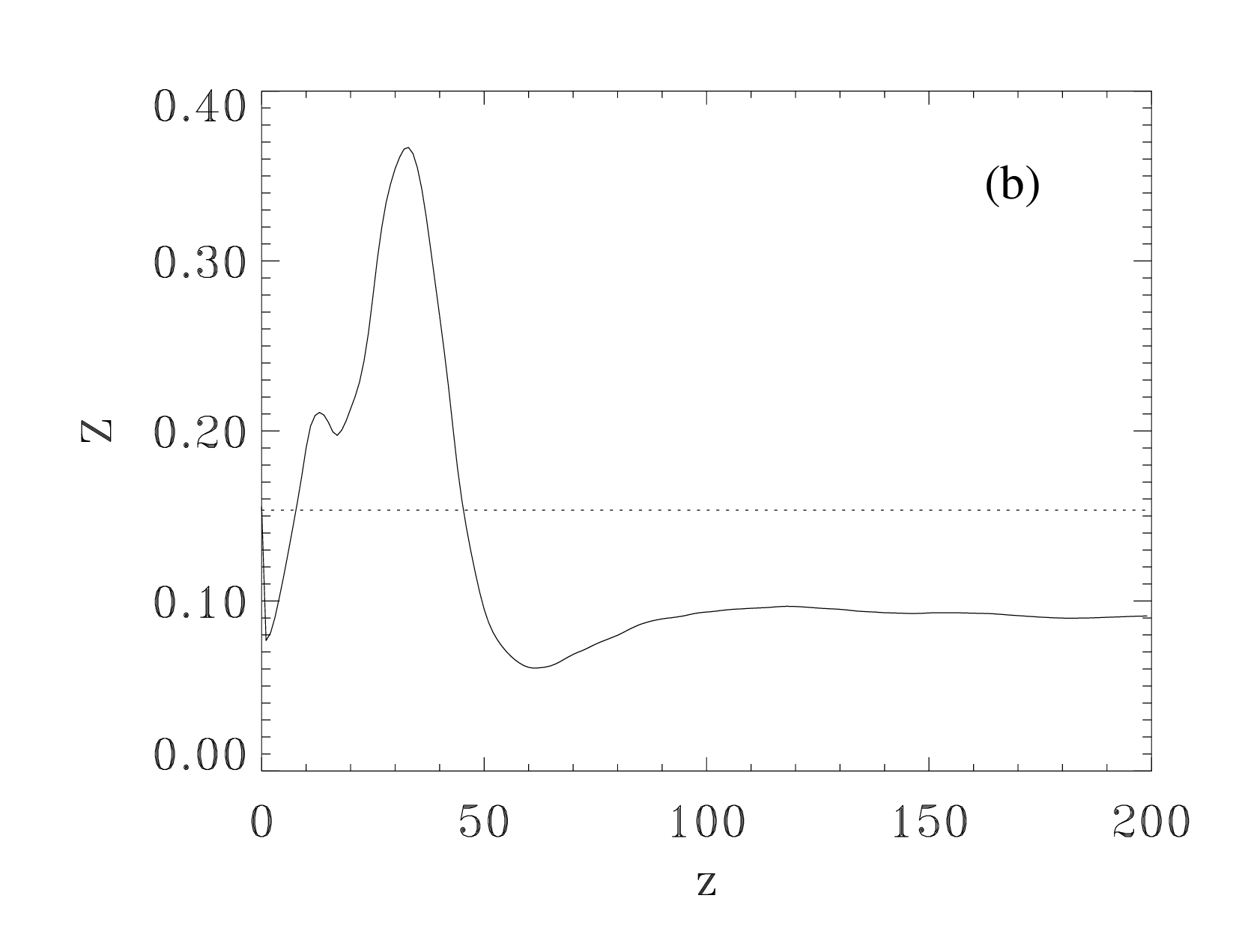}
\includegraphics{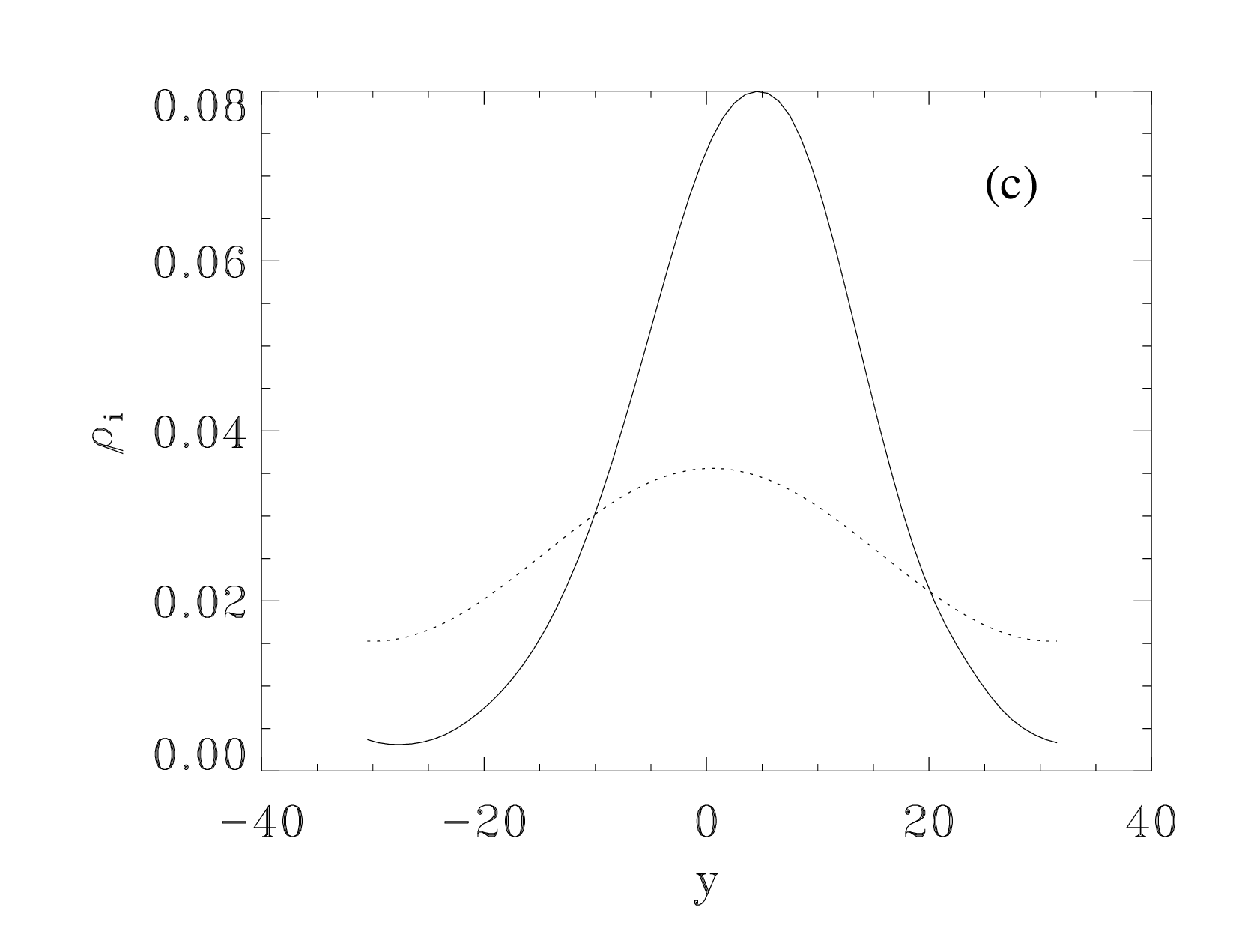}
\includegraphics{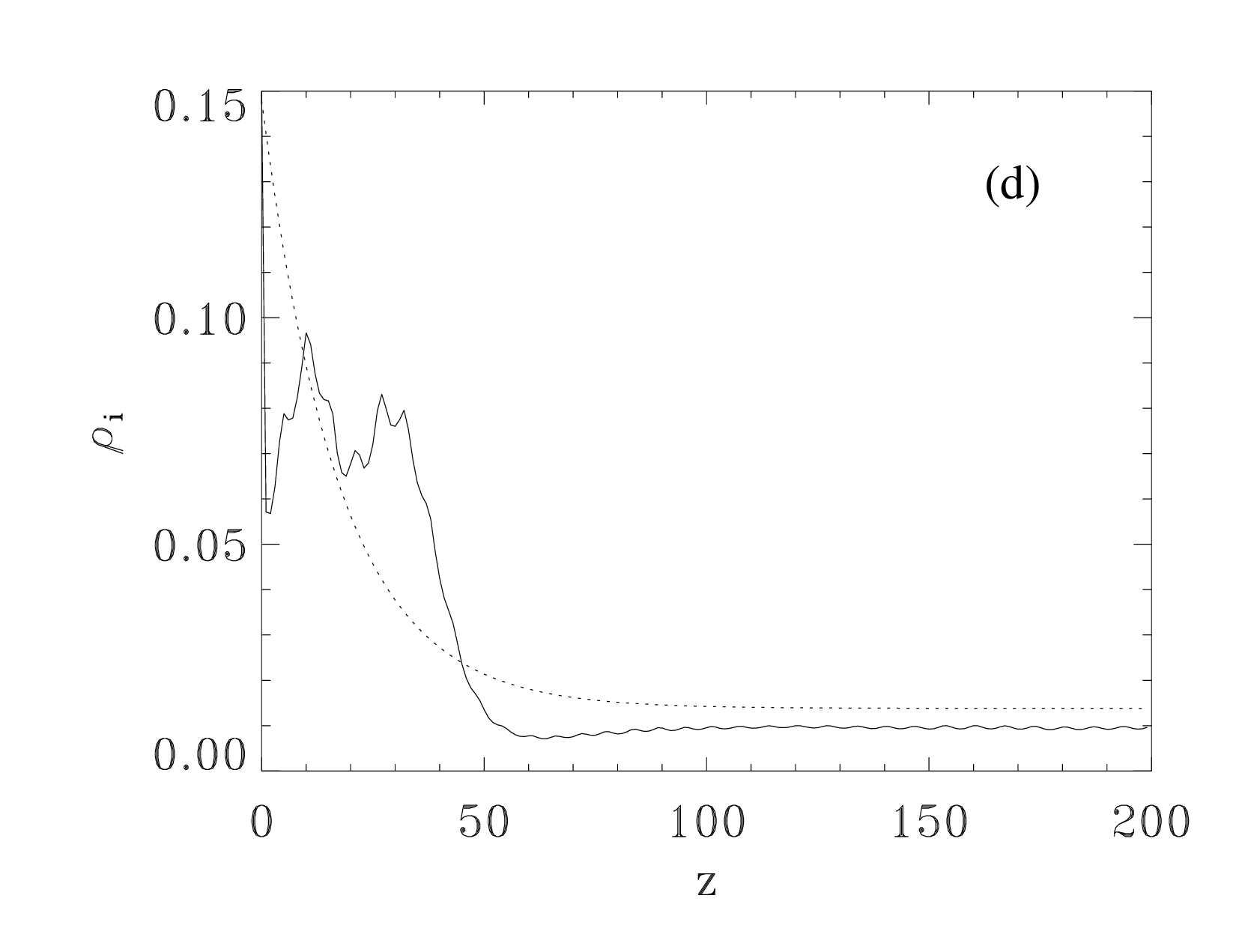}
\includegraphics{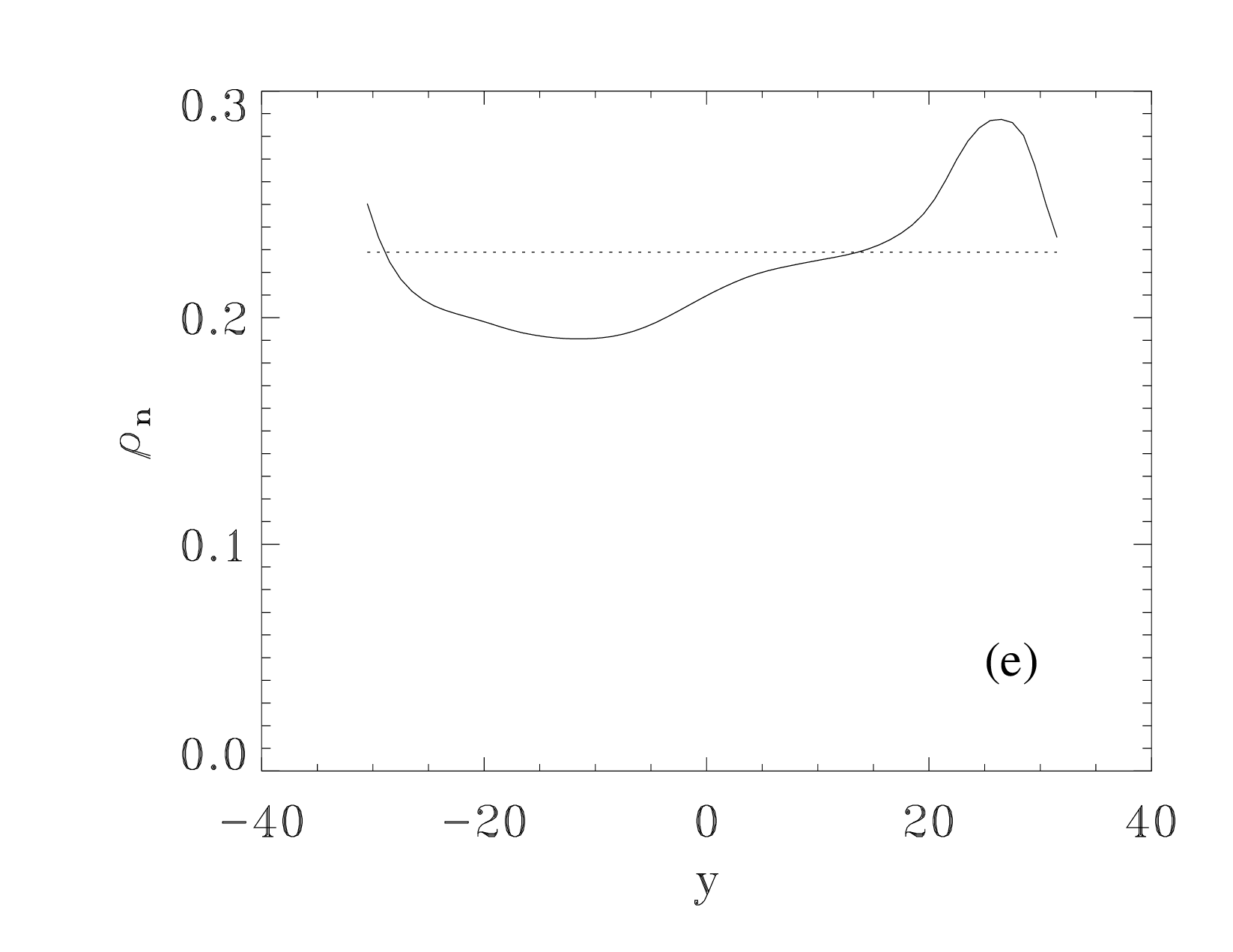}
\includegraphics{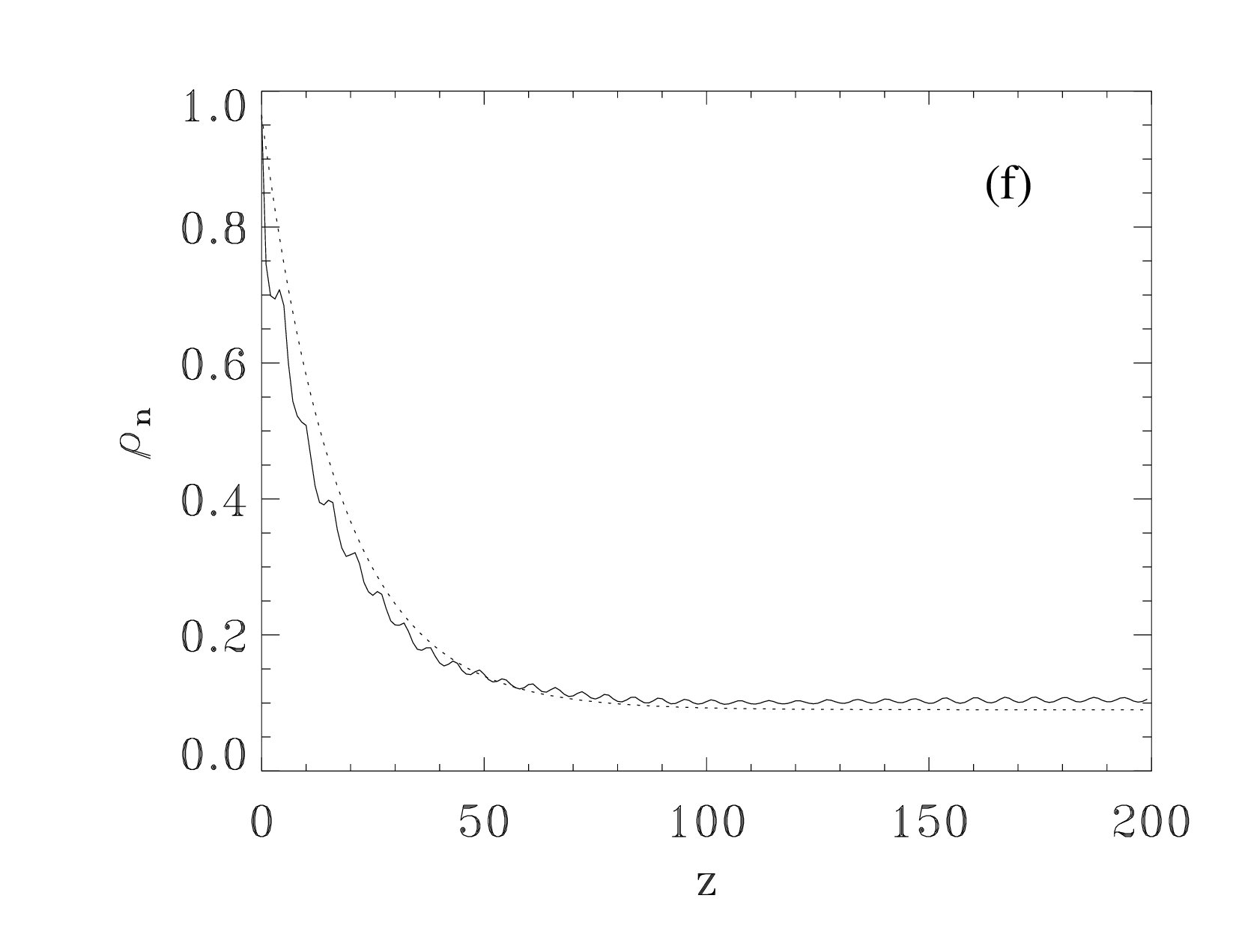}
\end{picture}
  \caption{ Top panels show the ionization fraction (left) ions density
(center)
and
neutrals density (right) profiles  in
$y$ direction at the altitude where the ionization fraction $Z$ reaches its
maximum.
Bottom panels are the respective vertical profiles at the same point. Dotted
lines
show the initial values.}
\label{3f}
\end{figure*}

However, at this stage the situation keeps evolving since the new
neutral and ions density profiles are more peaked, resulting in a
further peaking of $p_{Turb}$.  Thus, on a much longer time scale, the
ionization contrast will keep growing (\cite{tagger}); in our
simulations we find that typically $\sim 10^2$  Alfv\'en times
are needed to reach an equilibrium. Such long time scales mean that any
small numerical resistivity will result in a diffusion of the magnetic
flux, so that the filamentation of the magnetic field associated with
that of the ions cannot be observed.

\section{Results of the Numerical Experiments}

The simulations were carried out using the numerical model described in
the previous section.  They follow the evolution of a 2-D
($y$-horizontal and $z$-vertical directions), two fluid system (ions and
neutrals) with low ionization fraction.  The gas is initially threaded
by a vertical constant magnetic field and perturbed by horizontal waves
excited at the footpoints of the magnetic field lines ($z=0$).  Since no
approximations were made except the invariance in $x$, the following
parameters have to be imposed on each simulation: $B_0$, the initial
magnetic field, $c_s$, the sound velocity, $\mu$, the ion-neutral
collisional coefficient, $v_{t}$ and $\omega$, the amplitude and
frequency of the wave and the profiles of $\rho_i$ and $\rho_n$ in the
$y$ and $z$ directions.

In order to clarify the results presented in all figures, we must briefly
explain the scaling and units used by our numerical code. We have normalized
the parameters to the characteristic scales of the problem we are solving.
Therefore
we have taken as units the initial Alfv\'en velocity at $z=0$, $v_A$, and the
Alfv\'en time $t_A$, defined as the time that takes such an Alfv\'en wave in
crossing one wavelength  $\lambda_A$.

In the simulation presented in Figs. 1-6 we used a 61$\times$200 numerical
grid. It is longer in the vertical direction in order to allow the
complete damping of the wave.
As it was established in the previous section, we assume initial equilibrium
in a fluid  vertically
stratified and supported by gravity, with ions and neutrals
densities decreasing sharply (by a factor $\sim$ 10) with $z$
(shown in Fig. 1).
Initially the neutral density
is independent of $y$, but the ion density shows a small enhancement
in the $y$ direction, along which lie the perturbed velocities
associated with the wave. The magnetic field is adjusted so as to ensure MHD
equilibrium in the $y$ direction, given by (\ref{peq}).
The resulting initial ionization fraction $Z_0$ is constant over $z$ 
but shows a
maximum at $y=0$, as shown in Fig. 2 (left).

\begin{figure}[bh*]
\setlength{\unitlength}{1.0cm}
\begin{picture}(8,5.6)
\includegraphics{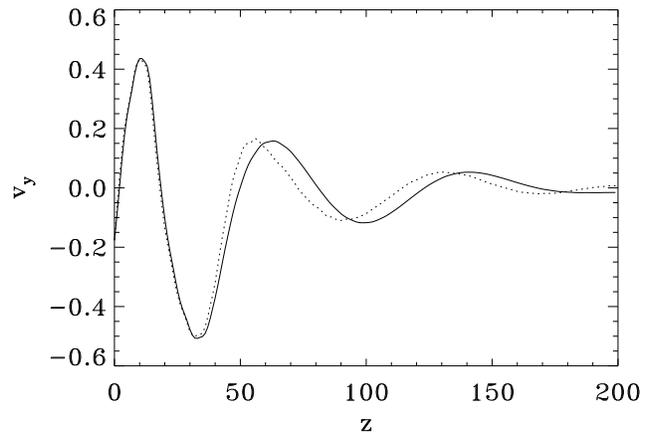}
\end{picture}
  \caption{ Vertical profile of the transverse perturbed velocities of 
ions (solid) and
neutrals (dotted). The decoupling of the two fluids begins at the altitude where
the wave is damped.}
\label{4f}
\end{figure}

A high amplitude Alfv\'en
wave is launched from $z=0$, by making
the whole fluid (neutrals, ions and magnetic
field lines) oscillate  in $y$ at a single frequency $\omega$;
this perturbation propagates upwards
as an Alfv\'en wave.
Its frequency is chosen so that the wave
propagates without damping at the lower part of the simulation grid
($\omega\ll
\nu_{ni}$), but is strongly damped at intermediate altitudes, where
$\rho_{i}$ (and thus $\nu_{ni}$) was taken to decrease sharply.
Therefore the
filamentation will occur only at the intermediate altitude where the wave
is damped (\cite{tagger}) but still retains a strong perturbed velocity.
In the simulation presented in Figs.  2-4 and 6, the perturbed velocity of the
wave at $z=0$ is of the order of the sound velocity 
($v_{t}=0.7c_{s}$), which is
taken to be a half of the Alfv\'en velocity.

\begin{figure}[bth*]
\setlength{\unitlength}{1.0cm}
\begin{picture}(8,5.6)
\includegraphics{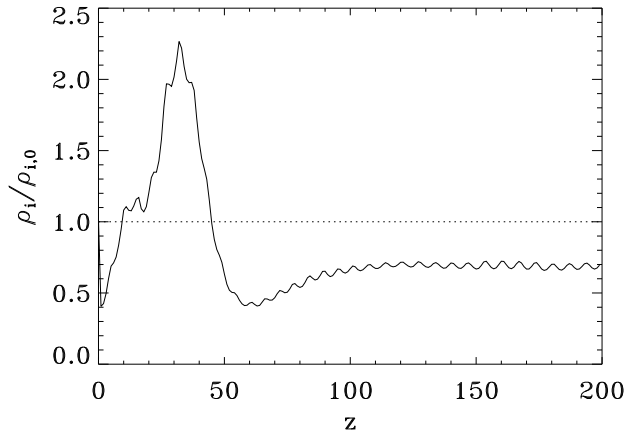}
\end{picture}
  \caption{ Vertical profile of the ratio between the final and 
initial ions density
at $y=y_{Z_{max}}$.}
\label{5f}
\end{figure}

Fig. 2 (right) shows the spatial profile of the ionization fraction  at the end
of the calculation. The initial contrast in the horizontal direction
has grown and shrunk  at a certain height.  At higher altitude
small traces of the wave still appear, together with a reversed 
profile along $y$.
An explanation to this behavior will be given later in this section.
In the vertical profile
of the ionization fraction (Fig. 3b) we note that the strong peak in 
Fig. 2 (right)
  is located
at $z\sim 32$, actually where the wave begins to be damped (see Fig. 4) but
still keeps enough amplitude to make the neutral expulsion mechanism effective.

Therefore we find a strong amplification
(by a factor $\sim 2.5$) of
the contrast in $Z$, at the altitude where the
wave is damped (Fig. 3a).
At the same altitude the ion density contrast along $y$, $\delta \rho_i$, grows
  and the more ionized region shrinks (Fig. 3c), as expected.
Moreover
the ion density decreases on the sides of the peak, suggesting ion 
motion towards the most
ionized regions.
In Fig. 5 we can  clearly quantify the final increase in ion density.
We also see in the simulation how the neutrals are expelled from the most
ionized regions (Fig. 3e), generating in the density profile a
'central' minimum of the same order as the increase achieved by the ions.
In all horizontal profiles of Fig. 3, the enhanced minimum/maximum are
not centered at $y=0$ because of the lateral motion due to the wave.
On Fig.~4 we compare the ion and neutral velocity. A phase lag
(corresponding to the damping of the wave by ion/neutral friction)
appears around $z=32$.

There is an additional effect acting to expel the neutrals from the most
ionized regions. Since
along those field lines the wave is less damped, its perturbed velocity at a
given height is larger.
However this contributes with a $v.\nabla v$ term on the ions as well as
on the neutrals, so that it should not contribute to the growth of the
ionization contrast.

Besides the strong peak in the spatial distribution of the ionization
fraction, there is another feature which is a direct output of the
enhancement of the ion density in the central magnetic field lines.  The
condensation of the ions must be accompanied by a compression of the
magnetic field lines at the same point.  Then magnetic tension causes
the compression of the field lines upwards.  But as noted before,
numerical resistivity has allowed the field to diffuse almost totally.  A
small residual intensification at the central field lines is left but it
is invisible at low altitudes, where the wave dominates the field
dynamics.  However, at the highest grid zones, where the wave is almost
completely damped, the increase in $B$ can be barely detected.  That
residual enhancement in $B$ causes the rise of the magnetic pressure,
which results in an expulsion of ions from the central field lines at
high altitudes.  This causes a reversed ion density profile, as observed
in Fig.  2.

In our simulations the filamentation process seems only limited by two
facts.  Firstly, $\rho_{i}$ can reach zero (so that the code crashes) at
large $y$.  In this case the process is so efficient that all the ions
at large $y$ are evacuated towards the most ionized regions, at the
altitude where the filamentation occurs.  \corr{Secondly, } $\rho_{i}$ can reach
zero in the lowest grid zones because the matter is pushed upwards by
the wave pressure.  Both processes impose severe restrictions to the
model parameters, in particular to the highest value of $v_{t}$ we can
use.  The second limitation could be easily overcome by changing the
boundary conditions to the more realistic case of allowing flux of
matter at $z=0$.  However, as discussed above, we prefer not to do it in
these first simulations since it would make it difficult to distinguish
the enhancements of ion density due to the filamentation process from
the ones due to this vertical flow.  This will be necessary, on the
other hand, in future realistic simulations of spicules.

\corr{Another unwanted effect is visible in our plots, for the largest
values of the turbulent velocities: perturbed quantities show a
small-scale sawtooth appearance, of numerical origin. They are
associated with sharp features in the vertical velocity, although this
component of the velocity remains much smaller than the horizontal one.
This phenomenon is generated at the first vertical grid points, and
results from a mismatch between the initial condition we impose ({\em
e.g.} equal ion and neutral velocities at $z=0$) and the wave
properties.  This effect remains small however and we have checked, by
varying the grid size, that it does not alter our results.}

Fig.  6 shows the evolution of the maximum value of the ionization
fraction for several initial wave amplitudes.  For the lowest values of
the perturbed velocity a state of equilibrium is achieved early in the
calculation.  However, for the highest velocities the filamentation is
more efficient and causes the simulation to crash.  In the simulation
shown in Fig.  3 ($v_{t}=0.35v_A$), this is due to the complete
depletion of ions at the first grid zones, although very small values of
$\rho_{i}$ are also achieved at high $y$, at the altitude where Z is
maximum.  In fact the curves corresponding to the highest velocities
show an oscillating  behavior over $t\sim 10^3t_A$, which can be a
consequence of the extremely low values of the ion density in those
regions.

Fig.  6 shows that the process acts faster for higher velocities.  The
asymptotic value of $Z_{max}$ varies quadratically with $v_t$, as
expected from the theory (\cite{tagger}).  However we find that this
value (and accordingly the horizontal pressure gradients obtained for
$\rho_i$ and $\rho_n$) depends on the initial profiles, {\it i.e.}, more
peaked initial profiles result in more peaked final ones.  Theory would
lead us to expect that the final equilibrium, balancing the
ponderomotive force of the wave with the pressure gradient of the
neutrals, should be independent of the initial state.  We believe that
this may be due to the vertical transport of ions and neutrals from the
lower and higher regions of the simulation grid.  In these regions the
filamentation process does not act, so that they retain a memory of the
initial conditions and can feed the region of wave absorption with
additional ions and neutrals.  However we have been unable to prove it
with the present limitations, due to the boundary condition at $z=0$.
We thus defer the treatment of this question to future work where the
detailed physics will be considered in more realistic conditions.

\begin{figure}[htb*]
\setlength{\unitlength}{1.0cm}
\begin{picture}(8,7)
\includegraphics{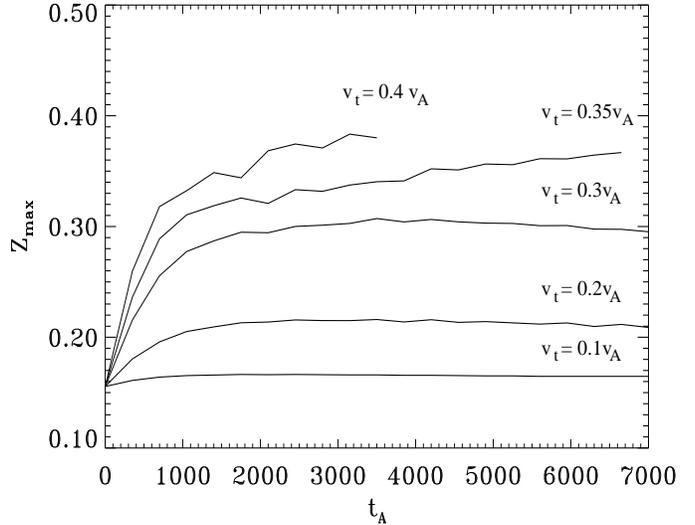}
\end{picture}
  \caption{ Time evolution of the value of maximum ionization fraction 
for different
initial perturbed velocities.}
\label{6f}
\end{figure}

\section{Discussion}
The aim of the simplified 2-D simulations presented in this paper was to
provide a first numerical test of the ambipolar filamentation mechanism.
We have tried to limit the physics involved to the minimal ingredients
needed, in order to clearly separate the expected physical effect (the
filamentation) from other sources of variation of ion and/or neutral
density.  In particular we have limited ourselves to the excitation of a
monochromatic wave and forbidden any inflow of matter from the lowest grid
boundary, although this turned out to cause the most
severe numerical limitation, by creating a vanishing ion density on the
first grid points.

We have found that the mechanism is strong and efficient, even with
these constraints and simplifying assumptions.  In the most efficient of
the simulations, the perturbed velocity was close to the sound velocity,
taken to be a half the Alfv\'en velocity (so the conditions are similar
to that of the interstellar medium).  However such velocities are still
lower than the observed supersonic motions which would make the
mechanism even more efficient, since we have found that its efficiency
goes with the square of the perturbed ion velocities.  We have obtained
similar results in cases closer to the conditions of the solar spicules,
where the sound velocity is a lower fraction of the Alfv\'en velocity
than in the results presented here ($c_{S}/v_{A}=.5$).

Future work will go in two directions: we will excite a more complete
spectrum of waves from the lowest boundary. This should cause the
filamentation to be less localized, since waves of different
frequencies will be damped (and cause filamentation) at different heights.

A more important evolution will be to relax the condition of no vertical
flow from $z=0$.  This will make the simulation more efficient since we
found that this boundary condition, chosen to clarify the physics,
caused the most severe numerical limitation.  It will also allow us to
make more realistic simulations of solar spicules.  In that case, we
expect our mechanism to act together with the vertical flow of matter
discussed in Haerendel (\cite{haerendel}), Tagger {\it et al} (1995) and
De Pontieu \& Haerendel (\cite{depon}).  Upward traveling Alfv\'en
waves generated in the photosphere are expected to cause both ambipolar
filamentation of the field lines and vertical acceleration of the gas,
so that a boundary condition allowing matter to flow vertically from
$z=0$ is necessary.  First tests actually show an intense growth of the
ionization fraction in the most ionized flux tubes.  We also expect
that, with these boundary conditions, we will be able to use higher
perturbed velocities making the mechanism much more efficient.

The effects of a more realistic equation of state and, of course, the
inclusion of ionization and recombination, are also obvious extensions to
this work. This should make us able to address more realistically the
role of our mechanism in the filamentary structure of the interstellar
medium and the star formation process. \corr{In particular, the simulations 
shown in this work correspond to ionization fractions similar to those 
observed in the warm neutral component of the Interstellar medium 
(\cite{kulkarni}). }

  Models of star formation
(since the early work of \cite{arons}) invoke
molecular clouds supported essentially by
turbulent magnetic pressure  and an ambipolar flow of
the neutrals toward dense cores (\cite{arons}; \cite{shu};
\cite{mckee}; \cite{vazquez}; and references therein).
In the
molecular clouds conditions, where the turbulence is supersonic, we
expect our mechanism to result in an additional pressure from the most
ionized to the less ionized regions that efficiently separates the ions
from the neutrals, favoring the gravitational collapse of the latter.

\corr{Another subject for future work could be the strong spatial and temporal
intermittency of MHD turbulence.  This is known to occur in intense
current and vorticity sheets (see \cite{spangler}, and references
therein; \cite{falga} and \cite{falga2}, for observations in the interstellar
medium and molecular
clouds).  We can expect our mechanism to be particularly efficient in
these sheets where gradients of all quantities become strong on
extremely small scales, and where turbulent motions concentrate.}

The turbulent magnetic fields can be generated as a consequence of the
Jeans-Parker instability during the molecular cloud formation process,
or (as suggested by recent models) by the outflow lobes associated with
young stellar objects; in the former case, large scale waves could feed
the turbulent cascade in the cloud (\cite{gomez}).
In the latter case, the outflows act as the sources of turbulence and
the energy released heats the cloud by the ambipolar drift (\cite{nomura}).
In those contexts, there would be localized
sources for the turbulence, that would  propagate in a stratified medium,
so that the physics studied here could be readily
applied.  On the other hand for a detailed study of the turbulent
cascade and its effects on the interstellar medium, a pseudo-spectral
technique would be better adapted to allow a random, non-localized
excitation of the waves.  The relevance of the role of the
ambipolar filamentation in those conditions will be the object of
future studies.

\begin{acknowledgements}
The authors wish to thank F. Masset and A. Hetem, who have been involved
in the early stages of the numerical developments used in this work, and
N. Hu\'elamo for the constructive discussions.  MF
has been supported by a Predoctoral Research Fellowship of the
Ministerio de Educaci\'on y Cultura (MEC) and by a C.I.E.S. (Centre
International des Etudiants et Stagiaires) grant. This work has been
partially supported by the MEC through the projects PB93-491 and PB97-269.

\end{acknowledgements}

\end{document}